\documentclass[conference]{IEEEtran}
\IEEEoverridecommandlockouts

\usepackage{graphicx}
\usepackage{subcaption}
\usepackage{cite}
\usepackage{amsmath,amssymb,amsfonts}
\usepackage{algorithmic}
\usepackage{graphicx}
\usepackage{textcomp}
\usepackage{xcolor}
\usepackage{booktabs}
\usepackage{tabularx} 
\usepackage{hyperref}
\usepackage{url}

\def\BibTeX{{\rm B\kern-.05em{\sc i\kern-.025em b}\kern-.08em
    T\kern-.1667em\lower.7ex\hbox{E}\kern-.125emX}}

\newcommand{\codeedu}{CodeEdu}

\DeclareRobustCommand*{\IEEEauthorrefmark}[1]{%
    \raisebox{0pt}[0pt][0pt]{\textsuperscript{\footnotesize\ensuremath{#1}}}}
\begin{document}

\title{CodeEdu: A Multi-Agent Collaborative Platform for Personalized Coding Education

 \thanks{*Equal contribution.}}

\author{\IEEEauthorblockN{Jianing Zhao\textsuperscript{*}\IEEEauthorrefmark{1},
Peng Gao\textsuperscript{*}\IEEEauthorrefmark{1},
Jiannong Cao\IEEEauthorrefmark{1},
Zhiyuan Wen\IEEEauthorrefmark{1}, 
Chen Chen\IEEEauthorrefmark{1},
Jianing Yin\IEEEauthorrefmark{1},
Ruosong Yang\IEEEauthorrefmark{2},
Bo Yuan\IEEEauthorrefmark{3}
}
\IEEEauthorblockA{\IEEEauthorrefmark{1}Department of Computing, The Hong Kong Polytechnic University, Hong Kong, China\\
 }
\IEEEauthorblockA{\IEEEauthorrefmark{2} China Mobile (Hong Kong) Innovation and Research Institute, Hong Kong, China}
\IEEEauthorblockA{\IEEEauthorrefmark{3}JIUTIAN Team, China Mobile Research Institute, Beijing, China\\
\{jianizhao, penggao, jiannong.cao, zhiyuan.wen, chen1.chen, jianing.yin\}@polyu.edu.hk \\
yangruosong@cmi.chinamobile.com, 
yuanboyjy@chinamobile.com}
}

\maketitle
\begin{abstract}
Large Language Models (LLMs) have demonstrated considerable potential in improving coding education by providing support for code writing, explanation, and debugging.
However, existing LLM-based approaches generally fail to assess students' abilities, design learning plans, provide personalized material aligned with individual learning goals, and enable interactive learning.
Current work mostly uses single LLM agents, which limits their ability to understand complex code repositories and schedule step-by-step tutoring.
Recent research has shown that multi-agent LLMs can collaborate to solve complicated problems in various domains like software engineering, but their potential in the field of education remains unexplored.
In this work, we introduce \codeedu, an innovative multi-agent collaborative platform that combines LLMs with tool use to provide proactive and personalized education in coding.
Unlike static pipelines, \codeedu~dynamically allocates agents and tasks to meet student needs.
Various agents in \codeedu~undertake certain functions specifically, including task planning, personalized material generation, real-time Q\&A, step-by-step tutoring, code execution, debugging, and learning report generation, facilitated with extensive external tools to improve task efficiency.
Automated evaluations reveal that \codeedu~substantially enhances students' coding performance.
A demonstration video of CodeEdu is available at \url{https://youtu.be/9iIVmTT4CVk}.

\end{abstract}

\section{Introduction}
Personalized education and intelligent tutoring are being enabled by artificial intelligence (AI) technology in education.
These systems are increasingly capable of designing learning plans, assessing learning progress, and providing personalized feedback~\cite{chu2503llm}.
However, these systems continue to struggle with understanding complex content~\cite{zhu2025recommender} and adapting to new knowledge domains, requiring fine-tuning for generalization.
\begin{figure}[ht!]
    \centering
    \includegraphics[width=0.9\linewidth]{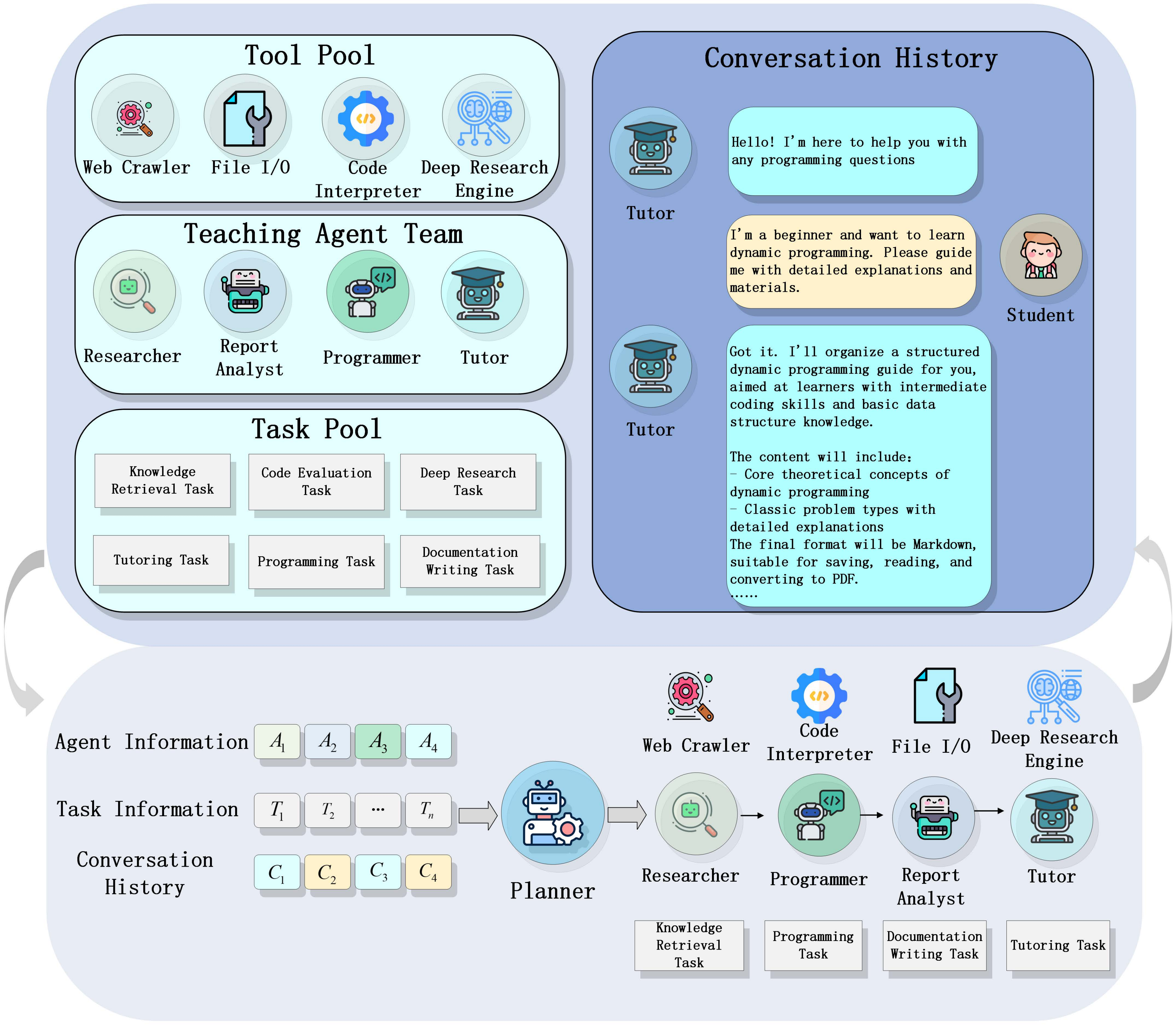}
    \caption{An overview of the \codeedu~platform. The figure is divided into three parts: (1) the top-left shows the system architecture and core modules; (2) the top-right shows the user interface; and (3) the bottom part illustrates the planning and collaboration process within agents.}
    \label{fig:framework}
\end{figure}
In recent years, the powerful natural language understanding of LLMs, along with the automation abilities of LLM agents, has introduced novel opportunities in the field of education, such as knowledge questions~\cite{tan2023can}, content summarization~\cite{zhang2024comprehensive}, and code generation~\cite{zhang2024codeagent}. 
In the field of code education, LLM agents can help students understand the logic behind coding languages~\cite{zhang2024codeagent, yang2024swe}.
In addition, they can also provide guidance during debugging~\cite{huang2024agentcodermultiagentbasedcodegeneration}.
However, teaching coding fundamentally differs from teaching static knowledge. 
It includes teaching students about abstract concepts, sequential implementation, code execution, and code evaluation.
Most current LLM-based tutoring solutions utilize a single-agent architecture, constraining their capacity to manage complicated workflows.
Specifically, single-agent systems struggle with multi-stage instructions, such as adapting to prior knowledge, scheduling exercises, debugging, and monitoring long-term learning objectives~\cite{tran2501multi}. 
Although individual LLMs can be augmented with memory or tool-calling functionalities, they are often reactive and cannot actively guide students through structured learning processes. 

By contrast, multi-agent systems (MAS) offer an advantageous framework for the decomposition and coordination of complex tasks. 
Specialized agents are assigned to various tasks like planning, content generation, and assessment~\cite{tran2501multi}.
Multi-agent LLM systems have recently shown advantages in software engineering, but their use in coding education remains underexplored.

In response, we propose \textbf{\codeedu}, a collaborative educational platform that leverages multi-LLM-powered agents for personalized, tool-enhanced coding education. 
In \codeedu, each agent plays a specific role tailored for distinct education tasks, such as planning, education, evaluating, or summarizing, enabling scalable and extensible learning pipelines. 
This platform encourages proactive planning and personalized learning, unlike passive Q\&A single-agent tutors.
Automated assessments reveal that \codeedu~enhances students' coding skills and provides high-quality learning materials. 

\section{Related Work}
\subsection{LLMs on Coding Education}
LLMs have become widely used in coding education in recent years.
LLM-based systems such as Codex~\footnote{\url{https://openai.com/zh-Hans-CN/index/introducing-codex/}}, Claude Coder\footnote{\url{https://www.anthropic.com/claude-code}}, and DeepSeek Coder~\cite{guo2024deepseekcoderlargelanguagemodel} can help students complete coding learning with simple instructions.
In code generation, CodeAgent~\cite{zhang2024codeagent} uses LLM-based agents with external tool integration to manage code repositories autonomously and adaptively.
Swe-Agent~\cite{yang2024swe} integrates LLMs into IDEs to enhance coding efficiency and simplify software development. 
In code education, AlgoBo~\cite{10.1145/3613904.3642349} introduces teachable LLM agents for code education, enhancing student knowledge in coding.
Single-agent architecture limits multi-step instruction, proactive guidance, and learner customization. 
\codeedu uses a collaborative multi-agent design to provide structured education, dynamic feedback, and personalized support, unlike single-LLM systems.

\subsection{Multi-agent Systems LLMs on Coding Education}
Recently, multi-agent systems have been utilized in coding education to improve interactivity and pedagogical efficacy. 
For instance, AgentCoder~\cite{huang2024agentcodermultiagentbasedcodegeneration} improves code generation quality by collaborating with coding and testing agents.
MapCoder~\cite{islam2024mapcodermultiagentcodegeneration} mimics human developer behaviors, and EduPlanner~\cite{zhang2025eduplanner} uses specialized agents for personalized curriculum design and optimization.
We provide a collaborative MAS in \codeedu, enabling personalized and adaptive learning.

\section{Framework}
\codeedu~is a multi-agent coding education platform driven by natural language input, aiming to provide flexible, scalable, and learner-centered intelligent education. The overall platform architecture is illustrated in Fig.~\ref{fig:framework}. The following sections present its overview, workflow, and implementation details.
\begin{figure*}[t]
    \centering
    \scalebox{0.8}{  
    \begin{minipage}{\textwidth}
        \centering
        \begin{subfigure}[t]{0.32\textwidth}
            \includegraphics[width=\linewidth]{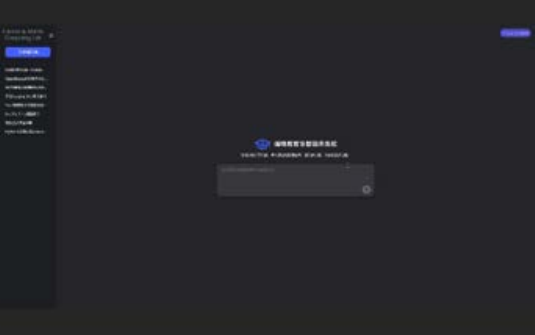}
            \caption{UI}
        \end{subfigure}
        \hfill
        \begin{subfigure}[t]{0.32\textwidth}
            \includegraphics[width=\linewidth]{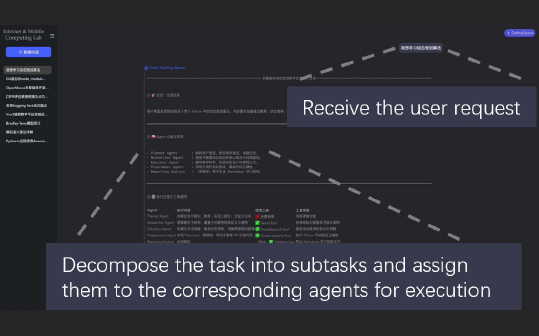}
            \caption{Task decomposition}
        \end{subfigure}
        \hfill
        \begin{subfigure}[t]{0.32\textwidth}
            \includegraphics[width=\linewidth]{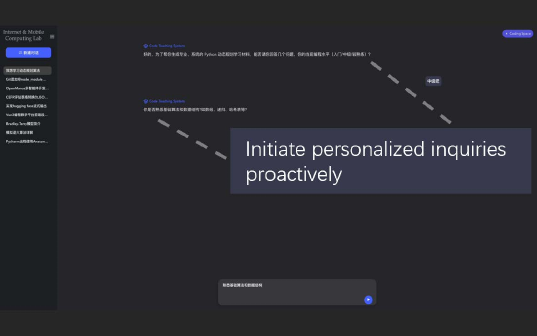}
            \caption{Personalized material generation}
        \end{subfigure}
    \end{minipage}
    }
    
    \scalebox{0.8}{
    \begin{minipage}{\textwidth}
        \centering
        \begin{subfigure}[t]{0.32\textwidth}
            \includegraphics[width=\linewidth]{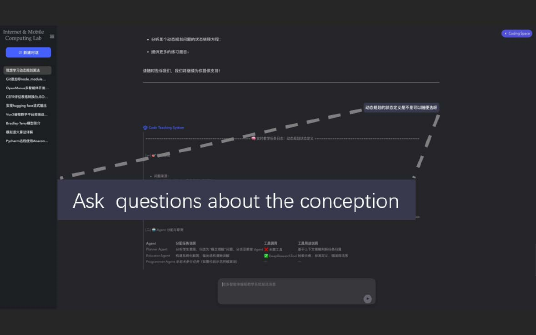}
            \caption{Real-time QA}
        \end{subfigure}
        \hfill
        \begin{subfigure}[t]{0.32\textwidth}
            \includegraphics[width=\linewidth]{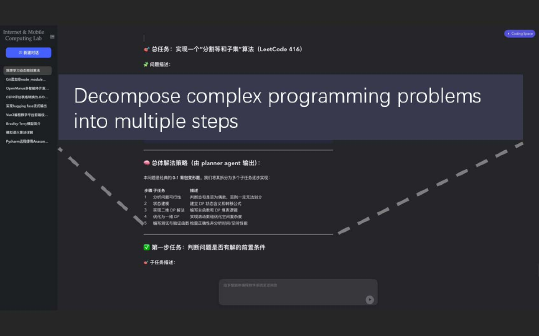}
            \caption{Step-by-step Code Tutoring with Debugging}
        \end{subfigure}
        \hfill
        \begin{subfigure}[t]{0.32\textwidth}
            \includegraphics[width=\linewidth]{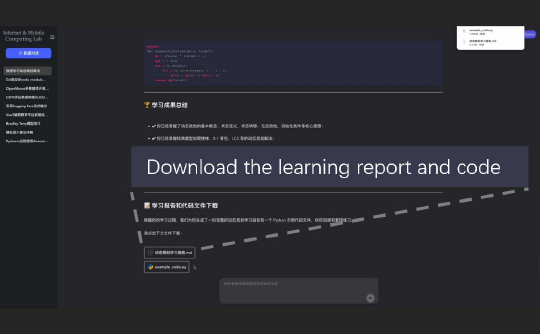}
            \caption{Learning report generation}
        \end{subfigure}
    \end{minipage}
    }
    
    \caption{A specific example of the workflow of \codeedu. (a) shows the user interface; (b) demonstrates task decomposition; and (c)--(f) correspond to the four core modules: Personalized Material Generation, Real-Time Q\&A, Step-by-step Code Tutoring with Debugging, and Learning Report Generation.}
    \label{fig:workflow}
\end{figure*}

\subsection{Overview}
CodeEdu is an interactive system. The UI is illustrated in Fig.~\ref{fig:framework} , with a detailed example provided in  Fig.~\ref{fig:workflow}. The system consists of three primary components: a \textit{Tool Pool}, an \textit{Agent Pool}, and a \textit{Task Pool} (shown in Fig.~\ref{fig:framework}). The \textit{Tool Pool} integrates common utilities, including a web crawler, file I/O, a code interpreter, and a deep research engine, providing foundational capabilities for various tasks. The \textit{Agent Pool} comprises five core agents with definitely defined roles: the \textit{Planner} decomposes and assigns tasks; the \textit{Researcher} retrieves external knowledge; the \textit{Report Analyst} records and summarizes the learning process; the \textit{Programmer} handles code execution and optimization; and the \textit{Tutor} offers real-time Q\&A. To ensure precise alignment between agents and tasks, the system provides the \textit{Task Pool}, consisting of six standard task types, including knowledge retrieval tasks, tutoring tasks, coding tasks, and so on. This classification assists the \textit{Planner} in efficient task scheduling. 
The system operates on an event-driven planner. 
Based on task information, agent information, and conversation history, the \textit{Planner} assigns the task to the most suitable agent for execution.

\subsection{Workflow}
The system has four core functions: Personalized Material Generation, Real-Time Q\&A, Step-by-step Code Tutoring with Debugging, and Learning Report Generation. The overall workflow is shown in Fig.~\ref{fig:workflow}.Each module may operate independently to satisfy specific learning objectives or be integrated into a cohesive instructional process. The ideal scenario is illustrated as follows.

Firstly, when receiving a learning request, the system conducts personalized inquiries (e.g., learning background) to create the user profile.  
The \textit{Planner} dynamically assigns tasks to the \textit{Researcher}, who uses the web crawler tool to gather and organize high-quality information based on user demands. 
This curated information is then compiled into \textbf{personalized learning materials}. 
As users study the material, they may encounter confusion with concepts or code examples.
Users can request clarification from the system at any time.
The \textit{Tutor} offers \textbf{real-time Q\&A} support and integrates learning material with internal knowledge to address user doubts.
After learning the material, next comes \textbf{step-by-step code tutoring with debugging}. 
The system can generate coding exercises aligned with the user's learning objectives. 
The \textit{Planner} will decompose the complex exercises into steps and provide prompts to guide the user to complete them. 
If the user submits code, the system will execute it and provide optimization or revision suggestions immediately.
Finally, the \textit{Report Analyst} compiles the learning trajectory, including user questions, code submissions, and system feedback, into a structured \textbf{learning report} for download and review.

\subsection{Implementation}
We implement \codeedu~ based on CrewAI~\footnote{\url{https://www.crewai.com/}}, a multi-agent platform that facilitates the rapid development of multi-agent systems for developers and provides a collection of external tools. 

In \codeedu, agent roles and task content are defined through manually designed prompts. 
Model APIs are manually selected for different types of agents based on functional alignment to satisfy their specific task requirements. 
For instance, the \textit{Programmer} is by default set by GPT-4o but can be flexibly replaced with more code-oriented LLMs, depending on the task requirements. 
What's more, the tools the system uses are provided by CrewAI. 
An overview of the tools and their functional descriptions is presented in Table~\ref{tab:table1}.

\begin{table}[htbp]
  \centering
  \caption{Illustrate the tools employed by each agent and provide a brief description of their functionalities.}
  \label{tab:table1}
  \resizebox{0.5\textwidth}{!}{
  \Large
  \begin{tabular}{>{\raggedright\arraybackslash}p{0.2\textwidth} 
                  >{\centering\arraybackslash}p{0.2\textwidth} 
                  >{\raggedright\arraybackslash}p{0.6\textwidth}}
    \toprule
    \textbf{Agent name} & \textbf{Tool name} & \textbf{Description of tools} \\
    \midrule
    Researcher    & Web Crawler & The tool is designed to search the internet and return the most relevant results.    \\ 
    Report analyst    & File IO     & The tool is desiged to read files from the local system and write content to fils.      \\ 
    Programmer   & Code Interpreter & The tool is designed for executing Python 3 code within a secure, isolated environment.     \\ 
    Tutor   & Deep Research Engine & The tool is designed to generate personalized explanations based on contextual information.      \\ 
    \bottomrule
  \end{tabular}
  }
\end{table}

\section{Experiment}
\subsection{Experiment Setup}
For automatic evaluation, we use LLMs to simulate students with varying levels of coding ability.
This study evaluates \codeedu's impact on learning improvement and material quality in comparison to conventional LLMs.

\subsubsection{Dataset}
We selected 100 LeetCode problems as the dataset, a popular platform in coding learning and evaluation. 
It covers a wide range of algorithmic subjects.
Standard input-output formats and unit test cases in each task ensure reliable automatic assessment of student-generated code.

\subsubsection{Baseline}
We use static prompting to configure GPT-4o as a coding tutor.

\subsubsection{Simulated Students}
We simulate students with three coding levels: a) \textbf{Low-level}: Receives only the LeetCode problem statement; b) \textbf{Medium-level}: Receives the problem and brief background concepts; and c) \textbf{High-level}: Includes the problem description, conceptual context, sample code, and optimized solution examples.
All students are created using GPT-4o.

We begin with a pre-test to assess students' coding ability. 
After that, each student receives education from \codeedu~or a baseline tutor agent via multi-turn chat on assigned LeetCode problems. 
Next, a post-test assesses their learning improvement.
The maximum number of dialogue turns is set to $T=20$, with early stopping determined by the LLM.
During the evaluation, each student can submit $k=3$ answers per problem, which will be assessed against $m=10$ unit use cases. 
We employ 5-fold cross-validation across $N=100$ coding problems.

\subsection{Evaluation Metrics}

We employ both $Pass@k$ and $Recall@k$ to assess coding level performance. 
These indicators allow us to evaluate both the correctness and completeness of code generated by students before and after education sessions.

\begin{itemize}
    \item \textbf{$Pass@k$}: measures the percentage of problems for which at least one of the top-$k$ generated code solutions passes all unit test cases.
    It reflects whether the student has learned to produce a fully correct solution.
    \begin{equation}
    \small
    \text{$Pass@k$} = \frac{1}{N} \sum_{n=1}^{N} \mathbf{1} \left[ \bigcup_{k=1}^{K} f_{p}(p_{nk}) \right]
    \label{eq:1}
    \small
\end{equation}
Where $N$ is the number of problems, $p_{nk}$ is the $k$-th code sample for problem $n$, and $f_p(\cdot)$ indicates whether the solution passes all test cases.

\item \textbf{$Recall@k$}: assesses the ratio of total test cases successfully passed among the top-$k$ answers. 
This metric assesses partial accuracy and the ability to address different edge cases.
\begin{equation}
\small
    \text{$Recall@k$} = \frac{1}{N \times M \times K} \sum_{n=1}^{N} \sum_{k=1}^{K} \sum_{m=1}^{M} \mathbf{1} \left[ f_r(p_{nk}, m) \right]
\small
\end{equation}
Where $M$ is the number of unit tests per problem, and $f_r(p_{nk}, m)$ checks whether the solution $p_{nk}$ passes the $m$-th test case.
\item \textbf{Tutor Improvement Rate (TIR)}: TIR evaluates the improvement of both $Pass@k$ and $Recall@k$, which can be formulated as:
\begin{equation}
\small
    \text{TIR}=\left(\frac{\mathcal{S}_{\text {post-test }}-\mathcal{S}_{\text {pre-test}}}{\mathcal{S}_{\text {pre-test }}}\right) \times 100 \%
\small
\end{equation}
$\mathcal{S}$ can be either $Pass$ or $Recall$.
\item \textbf{Evaluating the Quality of Learning Materials}:
We use GPT-4o to  evaluate the generated learning materials across four dimensions: Instructional Alignment (IA), Conceptual Clarity (CC), Interactivity (INT), and Personalization (PER), each rated on a 5-point Likert scale.
\end{itemize}

\subsection{Experimental Results}

\begin{figure}[h]
\centering
\includegraphics[width=0.46\textwidth]{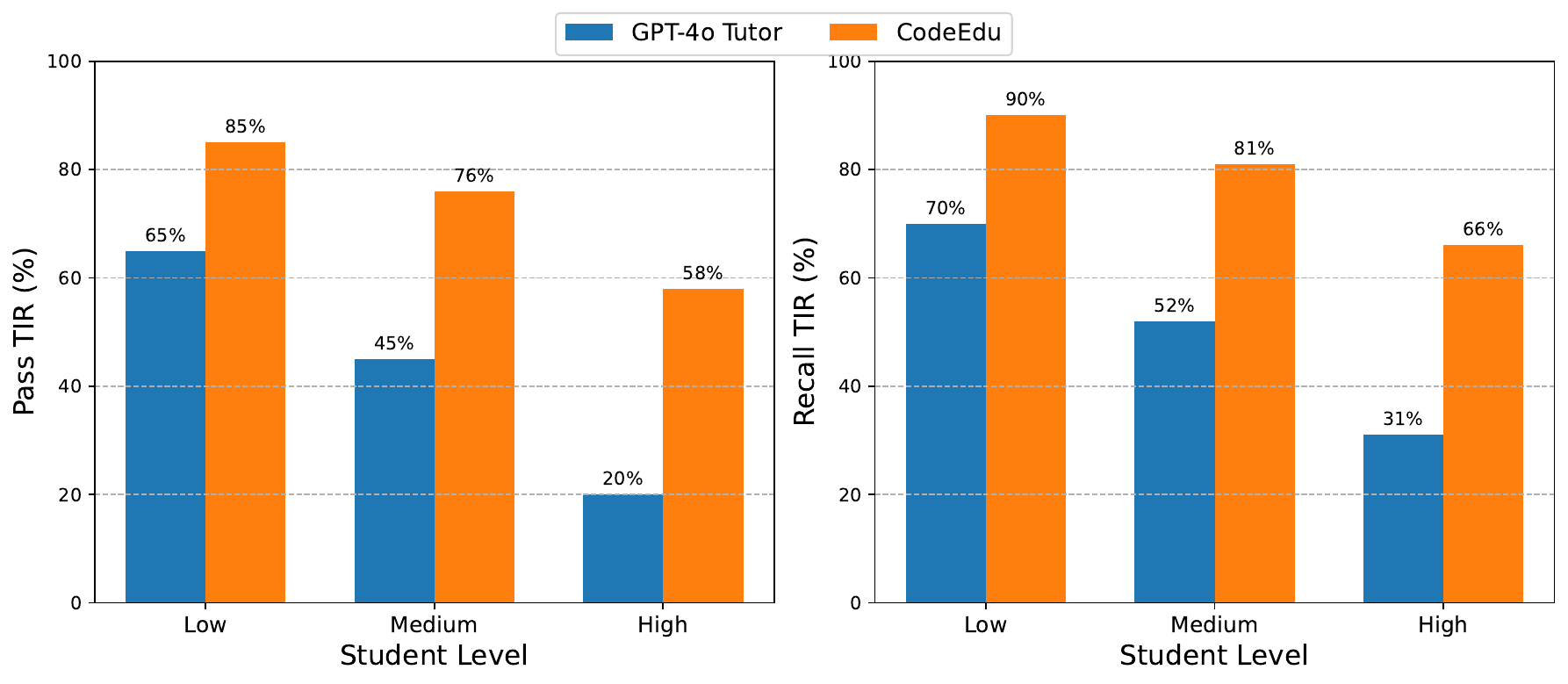}
\caption{Improvements in $Pass$ and $Recall$ scores across student levels using \codeedu~and the Baseline tutor.}
\label{fig:exe_1}
\end{figure}

As shown in Figure~\ref{fig:exe_1}, \codeedu~outperforms the baseline by 96.5\% in $Pass$ and 65.7\% in $Recall$. 
Both methods perform better for low and medium level students. 
Furthermore, for advanced students, \codeedu~markedly surpasses the baseline, achieving enhancements of 190\% and 113\% respectively, owing to its proactive support in guiding student learning, hence illustrating its efficacy in boosting coding ability.

\begin{figure}[h]
\centering
\includegraphics[width=0.25\textwidth]{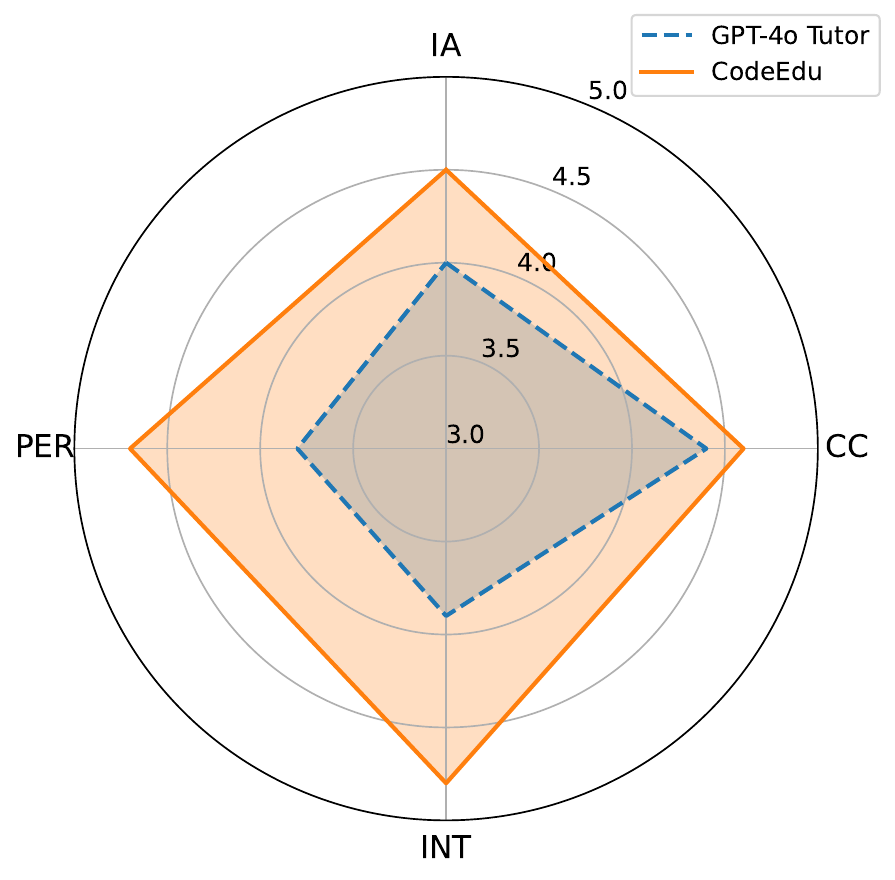}
\caption{Evaluate the quality of learning materials.}
\label{fig:exe_2}
\end{figure}

Figure~\ref{fig:exe_2} demonstrates that \codeedu~surpasses the baseline in the overall quality of learning materials by an average of 17.3\%. 
\codeedu~demonstrates significant enhancements in Interactivity (31.4\%) and Personalization (16.7\%), indicating improved learner engagement and adaptability.
These results show that \codeedu~is better at providing high-quality, personalized learning materials.

\section{Conclusion}
We introduce \codeedu, a multi-agent platform for coding education that facilitates structured, tool-enhanced, and personalized education. 
Automatic assessments indicate that it surpasses baseline in both educational outcomes and material quality. 
Future work includes the integration of human assessments, the facilitation of open-ended tasks, and the investigation of adaptive curricula for scalable personalization.

\end{document}